\documentclass[twocolumn,showpacs,preprintnumbers,amsmath,amssymb]{revtex4}
\usepackage[dvips]{graphicx}
\usepackage{dcolumn}
\usepackage{bm}

\begin{document}

\preprint{APS/123-QED}

\title{Lifetime Measurement of the ${}^3\!P_2$ Metastable State of Strontium Atoms}
\author{Masami Yasuda and Hidetoshi Katori}
\affiliation{%
Engineering Research Institute, 
The University of Tokyo, Bunkyo-ku, Tokyo 113-8656, Japan
}%

\date{\today}
\begin{abstract}
We have measured the lifetime of the $5s5p~{}^3\!P_2$ metastable state of 
strontium atoms by magneto-optically trapping the decayed atoms to the ground
 state, which allowed sensitive detection of the rare decay events.
 We found that the blackbody radiation-induced decay was the 
dominant decay channel for the state at $T=300$ K. The lifetime was 
determined to be $500^{+280}_{-130}$~s in the limit of zero temperature.
\end{abstract}

\maketitle

Precision measurements require the isolation of the physical system under 
study from environmental perturbations. The long coherence time thus obtained 
can be exploited to carry out ultra-high resolution spectroscopy \cite{Bergquist2000}
or to create and control the macroscopic quantum coherence in atomic systems,
 such as Bose-Einstein Condensate (BEC) \cite{BECNature} and Cooper pairing \cite{BCS}. 
Furthermore, the origin of environmental decoherence has recently attracted 
much attention in the context of quantum computation/communication, in which an 
entangled quantum system should maintain its internal and/or motional state 
coherence \cite{QC, Bloch}.
The lower-lying $^3\!P$ metastable states~\cite{KatoriICAP, Nagel, Hemmerich, 
Loftus, ErtmerMg, Udem, Wilpers} of alkaline earth species are intriguing candidates for these studies.
 A long metastable lifetime may allow an optical spectroscopy at the 1 mHz 
level ~\cite{Katori_Scottland}, enabling one to realize an ultra-precise atomic clock. 
In addition, the possibilities of evaporatively cooling them to reach BEC 
are discussed theoretically~\cite{SrBECDerevianko, SrBECKokoouline}. 

Laser-cooled and trapped atoms in ultra-high vacuum condition offer an ideal sample for these studies, 
as the collisional interactions with container walls or residual gases can be substantially removed. 
However, in some cases, the radiation from the surrounding walls dramatically affects the evolution 
of the internal state coherences.The room-temperature blackbody radiation~(BBR) has its intensity peak
 around a wavelength of $10\ \mu$m. Such BBR manifests itself most in Rydberg states  
\cite{Gallagher, Farley_Wing, Hollberg}, where infrared transitions with large electric dipole moments can be found.
However, because of large energy differences, it hardly excites atoms in the 
lower-lying states. Its influence, therefore, has been rarely discussed in laser cooling and trapping. 
On the other hand, it is known that the BBR causes small but non-negligible ac Stark shifts in ultra-precise 
spectroscopy~\cite{Itano,Bauch}. 

As for the ${}^3\!P$ metastable state of heavier alkaline-earth atoms, 
the upper-lying ${}^3\!D$ states are connected to the metastable state by 
electric dipole transitions with mid-infrared wavelength. Therefore,
 the excitation of the metastable state by the BBR may significantly alter
 its effective lifetime or introduce detectable blackbody shifts in precision spectroscopy.
In this Letter, we study the influence of the room-temperature BBR on the 
metastable state lifetime of Sr, which is recently predicted to be 1050~s \cite{Derevianko}. 
Because the lifetime is significantly longer than the collision-limited lifetime of tens of seconds that
 is realized in neutral atom traps at a vacuum pressure of
 $10^{-10}\ {\rm torr}$~\cite{Katori_Shimizu}, observing the survival of metastable atoms in such traps will
 not be practical~\cite{Katori_Shimizu,Ertmer}.
Instead of directly observing the decay of the metastable atoms by emitted photons~\cite{Walhout},
 we monitored the occurrence of the rare decay events to the ground state by magneto-optically 
trapping the atoms on the  ${}^1\!S_0 - {}^1\!P_1$ transition. In this way, we used the 
magneto-optical trap (MOT) as a photon amplifier with a gain of $10^{10}$ to detect 
the decay events with unit quantum efficiency~\cite{Nagourney,Bergquist}.

\begin{figure}
\begin{center}
	\scalebox{0.25}{\includegraphics{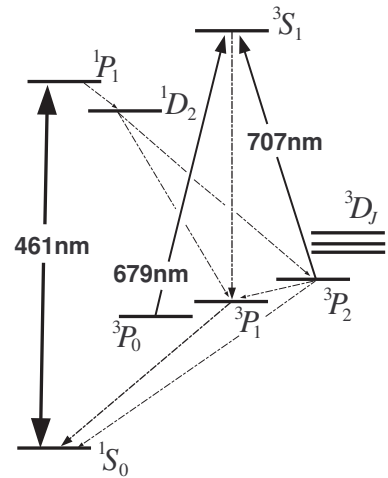}}
	\caption{The relevant energy levels for ${}^{88}\mathrm{Sr}$. The ${}^1\!S_0-{}^1\!P_1$ transition 
at $\lambda=461$~nm is used for trapping and detection.
The leak path through the ${}^1\!D_2$ state is used for loading atoms into the ${}^3\!P_2$ metastable state.
The ${}^3\!P_2 \rightarrow {}^3\!S_1$ transition at $\lambda=707$~nm and the ${}^3\!P_0 \rightarrow {}^3\!S_1$ transition at 
$\lambda=679$~nm are used to transfer the ${}^3\!P_2$ state population into the ground state. 
The upper-lying ${}^3\!D$ states are connected to the ${}^3\!P$ states by the electric dipole transitions at $\lambda \approx 3\ \mu$m.} 
\label{fig:EnergyLevel}
\end{center}
\end{figure}

Figure 1 shows the relevant energy levels for $^{88}$Sr.
Three radiative decay channels  from the metastable $5s5p\ {}^3\!P_2$ state 
have been identified by Derevianko~\cite{Derevianko}: 
1) The M2 transition to the ground state,  
2) Decay to the ${}^3\!P_1$ state through the M1, E2, and M3 transition,
3) Decay to the ${}^3\!P_0$ state through the E2 transition, where E$n$ or 
M$n$ stands for an electric or a magnetic $2^n$-pole transition.
The theory suggests that the contribution of the channel 3) is only 0.1\%~\cite{Derevianko}.
Therefore, 99.9\% of the population in the ${}^3\!P_2$ state finally relaxes to the ${}^1\!S_0$ ground state, 
as the atoms that decayed to the ${}^3\!P_1$ state further decay to the ${}^1\!S_0$ ground state in 
$22\ \mu{\rm s}$.
On this basis, we use the ${}^1\!S_0$ state population to detect the decay of the ${}^3\!P_2$ state. 

We measured the number of atoms $N_S(t)$ in the ${}^1\!S_0$ ground state by 
applying the MOT on the  ${}^1\!S_0-{}^1\!P_1$ transition.
The change of the atom number in the MOT is given by the rate equation,
\begin{equation}
	\frac{d}{dt}N_S(t)= -\left(\gamma_p+\Gamma_c\right) N_S(t)+\left(\gamma_r+\Gamma_q\right) N_P(t).
	\label{eqn:rateeqn}
\end{equation}
The atoms in the ground state are collisionally lost at a rate $\Gamma_c\approx (10~{\rm s})^{-1}$ 
or leaked out of the MOT transition via the $5s4d~{}^1\!D_2$ state with the typical 
rate of $\gamma_p\approx(20~{\rm ms})^{-1}$.
The atoms are supplied by the ${}^3\!P_2$ metastable state with its atom number of $N_P(t)$, 
which either radiatively decay ($\gamma_r$) or collisionally quench ($\Gamma_q$) to the ground state,
where $\Gamma_q \ll \gamma_r$ holds as discussed later. 
Ignoring both of the collisional loss terms of $\Gamma_c$ and $\Gamma_q$, 
the steady-state solution of Eq.~(1) for $t\gg \gamma_p^{-1}$ 
is  given by, 
\begin{equation}
	\gamma_r=\frac{N_S}{N_P}\gamma_p,
	\label{eqn:tau}
\end{equation}
where we assumed $N_P(t)$ to be constant as the decay of the population in the ${}^3\!P_2$ state 
with the collisional decay rate of $\Gamma_m\approx(10\ {\rm s})^{-1}$ is small enough in the time scale of interest.
The metastable decay rate $\gamma_r$, therefore, can be determined by the ratio $N_S/N_P$ of population 
distributed in both of the states and the MOT decay rate $\gamma_p$.
The number of atoms in the $^1\!S_0$ ground state can be derived by observing the MOT fluorescence intensity 
of $I_S=\eta N_S$, where $\eta$ is the photon counting rate per atom trapped in the MOT.
Similarly, $N_P$ can be obtained by transferring the metastable state population into the ground state 
and measuring the MOT fluorescence intensity of $I_P=\eta N_P$.
By the ratio of these fluorescence intensities, $\gamma_r/\gamma_p(=I_S/I_P)$ is accurately determined 
regardless of the coefficient $\eta$.

 The apparatus for magneto-optically trapping strontium atoms is similar to that described in Ref.~\cite{Katori}.
 The MOT fluorescence was collected by a lens with a solid angle of $10^{-5}$
 and then sent to a photomultiplier tube (PMT). 
An interference filter was placed in front of the PMT to block except the 461~nm light. 
 The output signal was then sent to a multi-channel scaler. 
The two transitions, ${}^3\!P_2 \rightarrow {}^3\!S_1$ and ${}^3\!P_0 \rightarrow {}^3\!S_1$ (Fig.~\ref{fig:EnergyLevel}), 
were used to pump the ${}^3\!P_2$ metastable state population into the ${}^1\!S_0$ ground state.

We first loaded the MOT from an atomic beam on the ${}^1\!S_0-{}^1\!P_1$ transition for 0.3~s. 
In the mean time, the ${}^3\!P_2$ metastable  state was populated via the weak branching 
decay to the ${}^1\!D_2$ state from the ${}^1\!P_1$ state, which is estimated to be $10^{-5}$ \cite{branch}.
About $10^7$ atoms in the low-field-seeking state were thus accumulated in the magnetic trap, 
which is formed by the quadrupole magnetic field used for the MOT.
The field gradient was 100 G/cm along its axis of symmetry.
At the typical peak density of $\approx10^{9}\ {\rm cm}^{-3}$, two-body collisional loss rate 
is estimated to be much smaller than the collisional loss term $\Gamma_c$ in Eq.~(\ref{eqn:rateeqn}), 
where we assumed an inelastic 
collisional loss rate of $\approx10^{-11}\ {\rm cm^3/s}$~\cite{SrBECDerevianko, SrBECKokoouline}.
We then turned off the MOT lasers and closed the shutter that blocked both the atomic beam and 
the thermal radiation from the oven.
We waited for 0.3~s so that all atoms except magnetically trapped metastable atoms diffused 
out of the trap region. 
After that, we turned on the MOT lasers again to capture atoms that were radiatively decayed 
from the ${}^3\!P_2$ state and recorded the fluorescence intensity.
At $t=1\ \mathrm{s}$, we irradiated both of the pumping lasers to transfer the metastable
 state population into the $^1\!S_0$ ground state and determined the number of atoms trapped 
in the ${}^3\!P_2$ state by the MOT fluorescence intensity.

\begin{figure}
\includegraphics[width=1.0\linewidth]{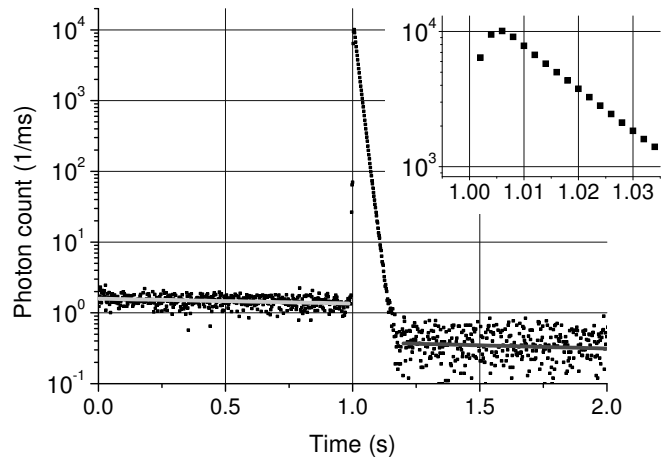}
\caption{The photon counting signal. The signal at $0<t<1$~s indicates the photons scattered by the 
recaptured atoms in the ground state.
 At $t=1~$s, the ${}^3\!P_2$ metastable state population is pumped into the ground state.
The peak intensity gives the number of atoms in the metastable state. The decaying slope gives 
the MOT lifetime, which is $\gamma_p^{-1}=$ 14.1~ms.
The inset shows the enlarged figure around the peak. The fluorescence intensity reaches its maximum in $\approx$ 5~ms.
}
\label{fig:mesresult}
\end{figure}

Figure~\ref{fig:mesresult}  shows a change of fluorescence intensity averaged over $10^2$ measurements, 
where the background level was subtracted by alternating the same procedure with 
and without loading atoms into the magnetic trap. 
The number of atoms in the ${}^1\!S_0$ state at $t=1$~s was determined by exponentially 
extrapolate the whole decay curve $I_S(t)$ of the MOT fluorescence to obtain $I_S(1)$ with better statistics.
The fluorescence decay in $0<t<1$~s was mainly caused by the collisional atom loss in the magnetic trap 
with the decay rate $\Gamma_m=(6.6\ {\rm s})^{-1}$ at the background gas pressure of $6.6 \times 10^{-10}\  {\rm torr}$.
By transferring the metastable state population into the ground state at $t=1\ \mathrm{s}$, the fluorescence 
intensity sharply rose up to its maximum  in about 5~ms. 
We approximated $I_P(1)$ by the peak fluorescence intensity shown in the inset of Fig.~\ref{fig:mesresult}.
The signal then decayed double-exponentially, consisting of the MOT decay with $\gamma_p=(14.1 \ {\rm ms})^{-1}$ 
due to the branching loss and the much slower collisional decay $\Gamma_m$ 
of the metastable atoms recaptured in the magnetic trap. 
The metastable state lifetime is calculated by applying the ratio $N_S/N_P=I_S(1)/I_P(1)$ 
and the measured MOT decay rate $\gamma_p$ in Eq.~(\ref{eqn:tau}). 
The measurement shown in Fig.~2 gave an effective radiative lifetime of $\gamma_r^{-1}=104^{+8}_{-7}$ s, 
which is only one tenth of the theoretical lifetime~\cite{Derevianko}. 
This shortening can be attributed to the BBR-induced decay via the $5s4d~{}^3\!D$ state as discussed later.
However, before discussing the BBR-induced decay, we checked the other decay channels.  

The reduction of the lifetime may be caused by the fine structure mixing of the metastable ${}^3\!P_2$ 
state with the ${}^3\!P_1$ or the ${}^1\!P_1$ state in the presence of the trapping magnetic field.
Assuming a magnetic field of 10~G, the magnetically induced decay rates are estimated to be 0.1\% of 
the natural decay rate $\gamma_0$ of the ${}^3\!P_2$ state. 
Actually, we measured the metastable lifetime under various magnetic field gradient, 
however, the change of the lifetime was within the statistical errors of 5 \% as expected. 
 Second, due to collisions with background gases, atoms in the metastable state may be 1) kicked out of 
the magnetic trap with the rate $\Gamma_m$ or 2) quenched into the ground state with $\Gamma_q$.
For the former issue, since we compared the fluorescence intensity $I_S$ and $I_P$ just before and 
after the population transfer, the collisional atom loss in the transferring period of $\tau_p=5$~ms may cause an error. 
However its fraction is estimated to be as small as $1-e^{-\Gamma_m \tau_p}\approx 10^{-3}$.
For the latter issue, the collisional quench to the ground state may cause pressure-dependent 
shortening of the metastable state lifetime.
To check this influence, we increased the background gas pressure up to $1.6\times 10^{-9}\ \mathrm{torr}$ 
and measured the lifetime. However, the change in the decay rate was well within statistical uncertainties. 

\begin{figure}
	\scalebox{0.8}{\includegraphics{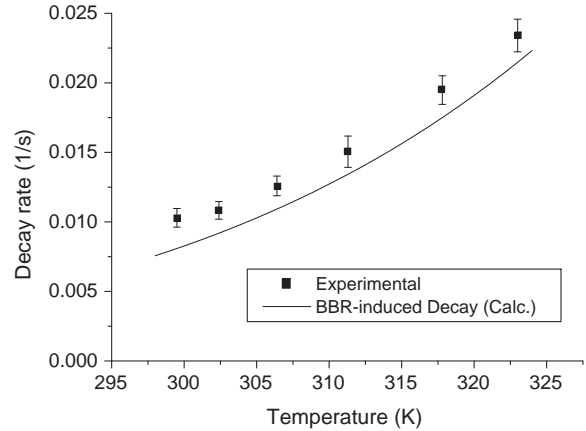}}
	\caption{Measured decay rate $\gamma_r$ under different ambient temperature $T$. 
The calculated BBR-induced decay rate $\gamma_B(T)$ 
is also shown. The offset between the measured data points and the calculated line 
indicates the natural decay rate $\gamma_0$.}
	\label{fig:DiffTemp}
\end{figure}

The rapid decay can be attributed to the metastable state quenching due to the BBR field 
that transfers atom population in the ${}^3\!P_2$ state to the short-lived ${}^3\!P_1$ state via the ${}^3\!D$ states.
By solving the coupled rate equations, the steady-state value of the BBR-induced decay rate $\gamma_B(T)$ is expressed as, 
\begin{equation}
	\gamma_B(T)=\gamma_D \frac{7}{36} \bar{n}(T),
\label{BBR}
\end{equation}
where $\gamma_D$ is the radiative decay rate of the $5s4d\,{}^3\!D$ state, 
and $\bar{n}(T)=\{\exp(h c/k_B T \lambda)-1\}^{-1}$ is the BBR photon occupation 
number at temperature $T$ with the wavelength $\lambda\approx 3.01\ {\rm \mu m}$ for the ${}^3\!P_2 - {}^3\!D $ transition. 
Using the radiative lifetime $\gamma_D^{-1}=2.9\pm0.2$~$\mu$s of the ${}^3\!D$ states \cite{Miller},
the BBR-induced decay rate is calculated to be $\gamma_{B}(T_0)= 8.03\times 10^{-3}\ \mathrm{s^{-1}}$ for $T_0=299.5$~K. 
Therefore the intrinsic decay rate $\gamma_0$ of the $^3\!P_2$ state  at 
$T \rightarrow  0\ $K is determined as 
$\gamma_0=\gamma_r(T_0)- \gamma_{B}(T_0) =(2.3\pm 0.7)\times 10^{-3}\  \mathrm{s^{-1}}$, 
which corresponds to a lifetime of $440^{+200}_{-100}$~s.
To confirm that this lifetime shortening originates in the BBR excitation, we measured the decay 
rate $\gamma_r(T)$ as a function of the ambient temperature $T$ by heating up the vacuum chamber that enclosed the MOT.
We note that this temperature change altered the vacuum pressure in the range of $(6.6-9.7)\times 10^{-10}\ {\rm torr}$.
However, the resultant collisional losses did not affect the measured lifetime as mentioned previously.
The measured decay rate $\gamma_r(T)$ is plotted in 
Fig.~\ref{fig:DiffTemp} by filled squares, where error bars indicate one standard deviation.
The monotonic increase of the decay rate as the temperature clearly supports that this lifetime 
shortening is caused by the BBR. 

The BBR-induced decay rate is estimated as follows.
Initially, the chamber that surrounds the atoms is assumed to be in thermal equilibrium.
However, in heating up the chamber, care should be taken if it is in thermal equilibrium. 
The vacuum chamber is made of stainless steel with 150 mm diameter and has six viewing ports 
made of BK7 glass with 40 mm diameter.
 It was heated by a ribbon heater wound around the chamber body.
We monitored the temperature of the viewing ports (at their rim and center) and 
the body of the chamber one hour after changing the heating power: Whereas the 
temperature of the rim was the same as that of the chamber body $T$, that of the center was somewhat lower by $\delta T(>0)$. 
Assuming the linear change of the temperature between the rim and the center, 
the average temperature for the viewing ports was approximated as $T_p= T-\delta T/3$.
Owing to this temperature inhomogeneity, atoms are not in the BBR field of thermal equilibrium.
Assuming a spherical radiant cavity, the effective photon occupation number 
for the body temperature of $T$ is calculated as~\cite{Chandos},
\begin{equation}
\bar{n}_{\mathrm{eff}}(T)=\frac{ \Omega_b \varepsilon_b \bar{n}(T)+ \Omega_p \varepsilon_p \bar{n}(T_p)} {\Omega_b \varepsilon_b+\Omega_p \varepsilon_p}.
	\label{eqn:neff}
\end{equation}
Here $\Omega_b$ is the solid angle covered by the chamber body and $\varepsilon_b$ 
is its spectral emissivity at $\lambda \approx 3\ \mu$m.
Likewise we defined $\Omega_p$ and $\varepsilon_p$ for the viewing ports, 
which fraction in the solid angle was $\Omega_p/\Omega_b \approx 10^{-1}$. 
We used this effective photon occupation number $\bar{n}_{\rm eff}(T)$ to 
calculate the BBR-induced decay rate given by Eq.~(\ref{BBR}). 
We took the spectral emissivity of both objects as fitting parameters 
so that the measured decay rate $\gamma_r(T)$ 
should have a constant offset $\gamma_0$ to the calculated BBR-induced decay rate $\gamma_B(T)$ at each temperature.
After the least squares fitting, we obtain the metastable decay rate 
of $\gamma_0=(2.1\pm0.4)\times 10^{-3}\ \mathrm{s^{-1}}$ and the spectral 
emissivities of $\varepsilon_b=1.6\%$ and $\varepsilon_p=88\%$ for stainless steel and BK7 glass, respectively. 
The obtained emissivity for BK7 glass showed fair agreement with $\varepsilon = 96\%$ 
derived from the index of refraction $n=1.48$ for BK7 glass at $\lambda \approx 3\ \mu$m. 
Reference data for polished stainless steel, however, was not available.

In the above discussion, we have not included the uncertainty of the ${}^3\!D$ state radiative lifetime, 
which gives coupling strength
 between the ${}^3\!P$ and ${}^3\!D$ states and thus significantly affects the BBR-induced decay rate.
  In Ref.~\cite{Miller}, the observed radiative lifetime of the ${}^3\!D$ state is given as $\tau_D=2.9 \pm 0.2\ \mu\mathrm{s}$. 
This uncertainty brings another statistical error of $\pm0.6\times 10^{-3}\ {\rm s}^{-1}$ to the metastable decay rate.
 Thus, we finally obtain the metastable decay rate $\gamma_0=(2.0 \pm 0.7) \times 10^{-3} \ \mathrm{s^{-1}}$, 
or the metastable lifetime of $500^{+280}_{-130}$~s. 

In summary, we have determined the lifetime of the $5s5p\ {}^3\!P_2$ state of $^{88}{\rm Sr}$ 
to be $500^{+280}_{-130}$~s in the limit of zero temperature.
Because the room-temperature BBR considerably shortens the metastable lifetime, 
care should be taken when dealing with the state to form a BEC \cite{SrBECDerevianko, SrBECKokoouline} or any other 
applications that require long coherence time. 
We, therefore, need to prepare cold environment to suppress thermal photons.
For example, by just lowering the ambient temperature down to 275~K, 
the BBR-induced decay rate becomes comparable to its intrinsic decay rate.
A straightforward comparison of the measured lifetime with the theory will be 
possible by further lowering the temperature down to 217~K, 
where the BBR quenching rate is expected to be 1\% of the natural decay rate.

The authors thank K. Okamura and M. Takamoto for their technical support. This work was  
supported by the Grant-in-Aid for Scientific Research (B) (12440110) from the Japan Society for the Promotion of Science.


\begin{thebibliography}{1}

	\bibitem{Bergquist2000} 
		R. J. Rafac {\it et al.},
		\prl
		\textbf{85},
		2462
		(2000).
	\bibitem{BECNature}
		J. R. Anglin and W. Ketterle,
		Nature
		\textbf{416},
		211-218
		(2002).
		See also references therein.
	\bibitem{BCS}
		K. M. O'Hara {\it et al.},
		Science
		\textbf{298},
		2179
		(2002).
	\bibitem{QC}
		C. Monroe {\it et al.},
		\prl
		\textbf{75},
		4714
		(1995).
	\bibitem{Bloch} 
		O. Mandel {\it et al.},
		http://arXiv.org/abs/cond-mat/0301169.
	\bibitem{KatoriICAP}
		H. Katori, T. Ido, Y. Isoya, and M. Kuwata-Gonokami,
		in \it Atomic Physics \rm
		\textbf{17},
		edited by E. Arimondo, P. DeNatale, and M. Inguscio
		(AIP, Melville, NY, 2001), p.~382.
	\bibitem{Nagel} 
		S. B. Nagel {\it et al.}, 
		\pra
		\textbf{67},
		011401(R)
		(2003).
	\bibitem{Hemmerich} 
		D. Hansen, J. Mohr, and A. Hemmerich,
		\pra
		\textbf{67},
		021401(R)
		(2003).
	\bibitem{Loftus} 
		T. Loftus, J. R. Bochinski, and T. W. Mossberg,
		\pra
		\textbf{66},
		013411
		(2002).
	\bibitem{ErtmerMg} 
		F. Ruschewitz {\it et al.},
		\prl
		\textbf{80},
		3173
		(1998).
	\bibitem{Wilpers} 
		G. Wilpers {\it et al.},
		\prl
		\textbf{89},
		230801
		(2002).
	\bibitem{Udem} 
		Th. Udem {\it et al.},
		\prl
		\textbf{86},
		4996
		(2001).
	\bibitem{Katori_Scottland}
		H. Katori,
		in \it Frequency Standards and Metrology, Proceedings of the Sixth Symposium\rm,
		edited by P. Gill
		(World Scientific, Singapore, 2002),
		p.~323.
	\bibitem{SrBECDerevianko}
		A. Derevianko {\it et al.}, 
		\prl
		\textbf{90},
		063002
		(2003).
	\bibitem{SrBECKokoouline}
		V. Kokoouline, R. Santra, and C. H. Greene,
		\prl
		\textbf{90},
		253201
		(2003).
	\bibitem{Gallagher} 
		T. F. Gallagher and W. E. Cooke,
		\prl
		\textbf{42},
		835
		(1979).
	\bibitem{Farley_Wing} 
		J. W. Farley and W. H. Wing,
		\pra
		\textbf{23},
		2397
		(1981).
	\bibitem{Hollberg} 
		L. Hollberg and J. L. Hall,
		\prl
		\textbf{53},
		230
		(1984).
	\bibitem{Itano} 
		W. M. Itano, L. L. Lewis, and D. J. Wineland,
		\pra
		\textbf{25},
		1233
		(1982).
	\bibitem{Bauch} 
		A. Bauch and R. Schr\"{o}der,
		\prl
		\textbf{78},
		622
		(1997).
	\bibitem{WalhoutXe} 
		M. Walhout, U. Sterr, A. Witte, and S. L. Rolston,
		Opt. Lett.
		\textbf{20},
		1192
		(1995).
	\bibitem{Xu} 
		X. Xu {\it et al.},
		J. Opt. Soc. Am. B
		\textbf{20},
		968
		(2003).
	\bibitem{Derevianko} 
		A. Derevianko,
		\prl
		\textbf{87},
		023002
		(2001).

	\bibitem{Katori_Shimizu} 
		H. Katori and F. Shimizu,
		\prl
		\textbf{70},
		3545
		(1993).
	\bibitem{Ertmer}
		M. Zinner {\it et al.}, 
		Phys.\ Rev.\ A
		\textbf{67},
		010501(R)
		(2003).
	\bibitem{Walhout} 
		M. Walhout, A. Witte, and S. L. Rolston,
		\prl
		\textbf{72},
		2843
		(1994).
	\bibitem{Nagourney} 
		W. Nagourney, J. Sandberg, and H. Dehmelt,
		\prl
		\textbf{56},
		2797
		(1986).
	\bibitem{Bergquist}
		J. C. Bergquist, R. G. Hulet, W. M. Itano, and D. J. Wineland,
		\prl
		\textbf{57},
		1699
		(1986).
	\bibitem{Katori} 
		H. Katori, T. Ido, Y. Isoya, and M. Kuwata-Gonokami,
		\prl
		\textbf{82},
		1116
		(1999).
	\bibitem{branch} 
		L. R. Hunter, W. A. Walker, and D. S. Weiss,
		\prl
		\textbf{56},
		823
		(1986).
	\bibitem{Miller} 
		D. A. Miller, L. You, J. Cooper, and A. Gallagher,
		\pra
		\textbf{46},
		1303
		(1992).
	\bibitem{Chandos} 
		R. J. Chandos and R. E. Chandos,
		Appl. Opt.
		\textbf{13},
		2142
		(1974).
	\end{thebibliography}
\end{document}